\documentclass[final,5p,times,twocolumn]{elsarticle}

\makeatletter
\def\@author#1{\g@addto@macro\elsauthors{\normalsize%
    \def\baselinestretch{1}%
    \upshape\authorsep#1\unskip\textsuperscript{%
      \ifx\@fnmark\@empty\else\unskip\sep\@fnmark\let\sep=,\fi
      \ifx\@corref\@empty\else\unskip\sep\@corref\let\sep=,\fi
      }%
    \def\authorsep{\unskip,\space}%
    \global\let\@fnmark\@empty
    \global\let\@corref\@empty  %% Added
    \global\let\sep\@empty}%
    \@eadauthor={#1}
}
\makeatother

\usepackage{amssymb}
\usepackage[section]{placeins} % This keeps floats inside their own section

\usepackage{upgreek}
\usepackage{xcolor}

\usepackage{amsmath}
\biboptions{numbers,sort&compress}
\newcommand{\electron}{\ensuremath{\mathrm{e}}}
\newcommand{\eplus}{\ensuremath{\mathrm{e}^{+}}}
\newcommand{\eminus}{\ensuremath{\mathrm{e}^{-}}}

\newcommand{\photon}{\ensuremath{\upgamma}}
\newcommand{\muon}{\ensuremath{\upmu}}
\newcommand{\muplus}{\ensuremath{\upmu^+}}
\newcommand{\muminus}{\ensuremath{\upmu^-}}

\newcommand{\piplus}{\ensuremath{\uppi^+}}
\newcommand{\piminus}{\ensuremath{\uppi^-}}

\newcommand{\uboson}{\ensuremath{\mathrm{U}}}

\newcommand{\Aprime}{\ensuremath{{\mathrm{A}^{\prime}}}}

\newcommand{\eV}{{e\kern-.06em V}}

\newcommand{\MeVc}{{\rm \,M\eV\kern-.05em /\kern-.02em c}}
\newcommand{\MeVcc}{{\rm \,M\eV\kern-.05em /\kern-.02em c$^{2}$}}

\newcommand{\GeVcc}{{\rm \,G\eV\kern-.05em /\kern-.02em c$^{2}$}}
\newcommand{\affuni}[2]{Dipartimento di Fisica dell'Universit\`a #1, #2, Italy.}
\newcommand{\affinfn}[2]{INFN Sezione di #1, #2, Italy.}
\journal{Physics Letters B}

\begin{document}

\begin{frontmatter}



%\title{Search for light vector boson production in $\eplus\eminus \to \uboson\upgamma$, $\uboson \to \eplus\eminus$ with the KLOE experiment}
\title{Limit on the production of a low-mass vector boson in $\eplus\eminus \to \uboson\upgamma$, $\uboson \to \eplus\eminus$ with the KLOE experiment}

%\collab{The KLOE-2 Collaboration}
\author[Messina,Frascati]{A.~Anastasi}
%\author[Roma2,INFNRoma2]{F.~Archilli},
\author[Frascati]{D.~Babusci}
%\author[Roma2,INFNRoma2]{D.~Badoni},
%\author[Cracow]{I.~Balwierz-Pytko},
\author[Frascati]{G.~Bencivenni}
\author[Warsaw]{M.~Berlowski}
%\author[Roma1,INFNRoma1]{C.~Bini},
\author[Frascati]{C.~Bloise}
%\author[INFNRoma1]{V.~Bocci},
\author[Frascati]{F.~Bossi}
\author[INFNRoma3]{P.~Branchini}
\author[Roma3,INFNRoma3]{A.~Budano}
%\author[Moscow]{S.~A.~Bulychjev},
\author[Uppsala]{L.~Caldeira~Balkest\aa hl}
\author[Uppsala]{B.~Cao}
%\author[Frascati]{P.~Campana},
%\author[Frascati]{G.~Capon},
\author[Roma3,INFNRoma3]{F.~Ceradini}
\author[Frascati]{P.~Ciambrone}
\author[Messina,INFNCatania,Novosibirsk]{F.~Curciarello}
\author[Cracow]{E.~Czerwi\'nski}
\author[Roma1,INFNRoma1]{G.~D'Agostini}
\author[Frascati]{E.~Dan\`e}
\author[INFNRoma3]{V.~De~Leo}
\author[Frascati]{E.~De~Lucia}
%\author[INFNBari]{G.~De~Robertis},
%\author[Roma1,INFNRoma1]{A.~De~Santis},
\author[Frascati]{A.~De~Santis}
%\author[Roma1,INFNRoma1]{G.~De~Zorzi},
\author[Frascati]{P.~De~Simone}
\author[Roma3,INFNRoma3]{A.~Di~Cicco}
\author[Roma1,INFNRoma1]{A.~Di~Domenico}
%\author[Napoli,INFNNapoli]{C.~Di~Donato},
\author[INFNRoma2]{R.~Di~Salvo}
%\author[Roma3,INFNRoma3]{B.~Di~Micco},
\author[Frascati]{D.~Domenici}
\author[Frascati]{A.~D'Uffizi}
%\author[Bari,INFNBari]{O.~Erriquez},
%\author[Bari,INFNBari]{G.~Fanizzi},
\author[Roma2,INFNRoma2]{A.~Fantini}
\author[Frascati]{G.~Felici}
\author[ENEACasaccia,INFNRoma1]{S.~Fiore}
%\author[Roma1,INFNRoma1]{P.~Franzini},
\author[Cracow]{A.~Gajos}
\author[Roma1,INFNRoma1]{P.~Gauzzi}
\author[Messina,INFNCatania]{G.~Giardina}
\author[Frascati]{S.~Giovannella}
%\author[Roma2,INFNRoma2]{F.~Gonnella},
\author[INFNRoma3]{E.~Graziani}
%\author[Cracow]{A.~Gruntowski} 
\author[Frascati]{F.~Happacher}
\author[Uppsala]{L.~Heijkenskj\"old}
%\author[Uppsala]{B.~H\"oistad},
%\author[Frascati]{L.~Iafolla},
%\author[Energetica,Frascati]{E.~Iarocci},
\author[Uppsala]{W.~Ikegami Andersson}
%\author[Uppsala]{M.~Jacewicz},
\author[Uppsala]{T.~Johansson}
%\author[Cracow]{K.~Kacprzak},
\author[Cracow]{D.~Kami\'nska}
\author[Warsaw]{W.~Krzemien}
%\author[Warsaw]{A.~Kowalewska},
%\author[Moscow]{V.~Kulikov},
\author[Uppsala]{A.~Kupsc}
%\author[Frascati,StonyBrook]{J.~Lee-Franzini},
%\author[Frascati]{B.~Leverington},
%\author[INFNBari]{F.~Loddo},
\author[Roma3,INFNRoma3]{S.~Loffredo}
\author[Messina,INFNMessina]{G.~Mandaglio}
%\author[Moscow]{M.~Martemianov},
\author[Frascati,Marconi]{M.~Martini}
\author[Frascati]{M.~Mascolo}
%\author[Moscow]{M.~Matsyuk},
\author[Roma2,INFNRoma2]{R.~Messi}
\author[Frascati]{S.~Miscetti}
\author[Frascati]{G.~Morello}
\author[INFNRoma2]{D.~Moricciani}
\author[Cracow]{P.~Moskal}
%\author[INFNRoma3,LIP]{F.~Nguyen},
%\author[Frascati]{L.~Quintieri}
\author[Frascati]{A.~Palladino\corref{cor1}\fnref{fn1}}
\fntext[fn1]{Present address: Department of Physics, Boston University, Boston, USA}
\ead{palladin@bu.edu}
\author[Uppsala]{M.~Papenbrock}
\author[INFNRoma3]{A.~Passeri}
\author[Energetica,INFNRoma1]{V.~Patera}
\author[Frascati]{E.~Perez~del~Rio}
%\author[Roma3,INFNRoma3]{I.~Prado~Longhi},
\author[INFNBari]{A.~Ranieri}
%\author[Mainz]{C.~F.~Redmer},
\author[Frascati]{P.~Santangelo}
\author[Frascati]{I.~Sarra}
\author[Calabria,INFNCalabria]{M.~Schioppa}
%\author[Frascati]{B.~Sciascia},
%\author[Energetica,Frascati]{A.~Sciubba},
\author[Frascati]{M.~Silarski}
\author[Frascati]{F.~Sirghi}
%\author[Calabria,INFNCalabria]{S.~Stucci},
%\author[Roma3,INFNRoma3]{C.~Taccini},
\author[INFNRoma3]{L.~Tortora}
\author[Frascati]{G.~Venanzoni\corref{cor1}}
\ead{graziano.venanzoni@lnf.infn.it}
%\author[Frascati,CERN]{R.~Versaci},
\author[Warsaw]{W.~Wi\'slicki}
\author[Uppsala]{M.~Wolke}
%\author[Cracow]{J.~Zdebik}
%%%%%%%%%%%%%%%%%%%%%%%%%%%%%%%%%%%%%%%%%%%%%%%%%%%%%%%%%%%%%%%%%%%%%%%%%%%%%%%%%%%%%%%%%%%%%%%%%%
%\address[Bari]{\affuni{di Bari}{Bari}}
\address[INFNBari]{\affinfn{Bari}{Bari}}
%\address[CentroCatania]{Centro Siciliano di Fisica Nucleare e Struttura della Materia, Catania, Italy.}
\address[INFNCatania]{\affinfn{Catania}{Catania}}
\address[Cracow]{Institute of Physics, Jagiellonian University, Cracow, Poland.}
\address[Frascati]{Laboratori Nazionali di Frascati dell'INFN, Frascati, Italy.}
%\address[Messina]{\affuni{di Messina}{Messina}}
%\address[Mainz]{Institut f\"ur Kernphysik, 
%Johannes Gutenberg Universit\"at Mainz, Germany.}
\address[Messina]{Dipartimento di Fisica e Scienze della Terra dell'Universit\`a di Messina, Messina, Italy.}
\address[INFNMessina]{INFN Gruppo collegato di Messina, Messina, Italy.}
\address[Calabria]{\affuni{della Calabria}{Rende}}
\address[INFNCalabria]{INFN Gruppo collegato di Cosenza, Rende, Italy.}
%
%\address[Moscow]{Institute for Theoretical and Experimental Physics (ITEP), Moscow, Russia.}
%\address[Napoli]{\affuni{``Federico II''}{Napoli}}
%\address[INFNNapoli]{\affinfn{Napoli}{Napoli}}
\address[Energetica]{Dipartimento di Scienze di Base ed Applicate per l'Ingegneria dell'Universit\`a 
``Sapienza'', Roma, Italy.}
\address[Marconi]{Dipartimento di Scienze e Tecnologie applicate, Universit\`a ``Guglielmo Marconi", Roma, Italy.}
\address[Novosibirsk]{Novosibirsk State University, 630090 Novosibirsk, Russia.}
\address[Roma1]{\affuni{``Sapienza''}{Roma}}
\address[INFNRoma1]{\affinfn{Roma}{Roma}}
\address[Roma2]{\affuni{``Tor Vergata''}{Roma}}
\address[INFNRoma2]{\affinfn{Roma Tor Vergata}{Roma}}
\address[Roma3]{Dipartimento di Matematica e Fisica dell'Universit\`a 
``Roma Tre'', Roma, Italy.}
%\address[Roma3]{\affuni{``Roma Tre''}{Roma}}
\address[INFNRoma3]{\affinfn{Roma Tre}{Roma}}
\address[ENEACasaccia]{ENEA UTTMAT-IRR, Casaccia R.C., Roma, Italy}
%\address[StonyBrook]{Physics Department, State University of New 
%York at Stony Brook, USA.}
\address[Uppsala]{Department of Physics and Astronomy, Uppsala University, Uppsala, Sweden.}
\address[Warsaw]{National Centre for Nuclear Research, Warsaw, Poland.}
%\address[CERN]{Present Address: CERN, CH-1211 Geneva 23, Switzerland.}
%\address[LIP]{Present Address: Laborat\'orio de Instrumenta\c{c}\~{a}o e F\'isica Experimental de Part\'iculas,
%Lisbon, Portugal.}
%\address[BU]{Department of Physics, Boston University, Boston, USA}

\cortext[cor1]{Corresponding authors}

%\address{}

\begin{abstract}
The existence of a new force beyond the Standard Model is compelling 
because it could explain several striking astrophysical observations which fail 
standard interpretations. We searched for the light vector mediator of 
this dark force, the \uboson~boson, with the KLOE detector at the DA$\Phi$NE \eplus\eminus{} collider. 
Using an integrated luminosity of 1.54 fb$^{-1}$, we 
studied the process $\eplus\eminus \to \uboson\photon$, 
with $\uboson \to \eplus\eminus$, using radiative return to search for a resonant peak in the dielectron 
invariant-mass distribution. We did not find evidence for a signal, and 
set a 90\%~CL upper limit on the mixing strength between 
the Standard Model photon and the dark photon, $\varepsilon^2$, at $10^{-6}$--$10^{-4}$ in 
the 5--520~MeV/c$^2$ mass range. 
\end{abstract}

\begin{keyword}
dark matter \sep dark forces \sep dark photon \sep \uboson~boson \sep \Aprime


\end{keyword}

\end{frontmatter}


\section{Introduction}
\label{sec:intro}
The Standard Model (SM) of particle physics has received further confirmation with the discovery of the 
Higgs boson~\cite{higgs1,higgs2,higgs3}, however, there are strong hints of physics it cannot explain, such as 
neutrino oscillations~\cite{review_nu_oscillations_PDG} and the measured anomalous magnetic moment of the muon~\cite{review_amu}. %~\cite{Agashe:2014kda}. 
Furthermore, the SM does not provide a dark matter (DM) candidate usually 
advocated as an explanation of the numerous gravitational anomalies observed 
in the universe. Many extensions of 
the SM~\cite{pospelovPLB2008,undici,dodici,tredici,diciannove} consider a 
Weakly Interacting Massive Particle (WIMP) as a viable
DM candidate and assume that WIMPs are charged under
a new kind of interaction. The mediator of the new force would be a
gauge vector boson, the \uboson{}~boson, also referred to as a dark photon 
or \Aprime{}. It would be produced during WIMP annihilations, have a mass 
less than two proton masses, and a leptonic decay channel in order to 
explain the astrophysical observations recently reported by 
many experiments~\cite{integral,pamela,ams,atic,fermi,hess1,hess2,dama1,dama2,cogent1,cogent2}. 
%Although there are alternative explanations for some of these anomalies, they could 

In the minimal theoretical model, the \uboson~boson is the lightest particle 
of the dark sector and can couple to the ordinary SM photon only through loops 
of heavy dark particles charged under both SM~U(1)$_{\rm Y}$ and 
dark U(1)$_{\rm D}$ symmetries~\cite{darksector1,darksector2,drees2006,darksector3,darksector4,pospelovPLB2008}. 
These higher-order interactions would open a so-called kinetic mixing portal described in the theory by 
the Lagrangian term   
%$\mathcal{L}_{\mathrm{mix}} = -\,{}^{\varepsilon}\!\!/_{2} \, F^{\mathrm{EW}}_{ij}F^{ij}_{\mathrm{Dark}}$, 
$\mathcal{L}_{\mathrm{mix}} = -\,\frac{\varepsilon}{2}\, F^{\mathrm{EW}}_{ij}F^{ij}_{\mathrm{Dark}}$, 
where $F^{\mathrm{EW}}_{ij}$ is the SM hypercharge gauge field tensor 
and $F^{ij}_{\mathrm{Dark}}$ is the dark field tensor. The $\varepsilon$ 
parameter represents the mixing strength and is the ratio of the dark and 
electromagnetic coupling constants. In principle, the dark photon could be 
produced in any process in which a virtual or real photon is involved but 
the rate is suppressed due to the very small coupling 
%it would be a very rare event being the size of the coupling very small 
($\varepsilon < 10^{-2}$). 
In this respect, high-luminosity $\mathcal{O}$(GeV)-energy \eplus{}\eminus{}
colliders play a crucial role in dark photon searches~\cite{venti,ventuno,ventidue}. 

We investigated the $\eplus\eminus \to \uboson \photon$  
process by considering the \uboson~boson decaying into $\eplus \eminus$. 
At the level of coupling accessible by KLOE in this channel the \uboson~boson is expected to decay promplty leaving its signal as a
resonant peak in the invariant-mass distribution of the lepton pair. 
The energy scan was performed by applying the radiative return method which consists of  
selecting the events in which either electron or positron emits an initial-state radiation (ISR) photon which carries away
a part of the energy and effectively changes the amount of the energy available for \uboson~boson production.
%At the level of coupling accessible by KLOE in this channel the \uboson~boson is expected to decay promptly leaving its signal as a 
%resonant peak in the invariant-mass distribution of the lepton pair, scanned by the hard 
%initial-state radiation (ISR) photon's radiative return. 
%In our analysis the selected $\photon$ originates from ISR allowing 
%us to increase the sensitivity of $\uboson \to \eplus\eminus$ decay. 
The selected initial- and final-state particles are the same as in the 
radiative Bhabha scattering process so 
we receive contributions from resonant $s$-channel, non-resonant $t$-channel
\uboson~boson exchanges, and from $s$-$t$~interference. The finite-width effects 
related to $s$-channel annihilation sub-processes, scattering $t$-channel and
$s$-$t$~interference are of order of $\Gamma_{\uboson}/m_{\uboson}$ for the 
integrated cross section and can be neglected with respect to any potential 
resonance we would observe; $\Gamma_{\uboson} \sim 10^{-7}$--$10^{-2}$~MeV for the 
coupling strengths to which we are sensitive~\cite{barze_2011}. The non-resonant $t$-channel effects would not 
produce a peak in the invariant-mass distribution but could, 
in principle, appear in analyses of angular distributions or asymmetries. 
We are going to report exclusively on resonant $s$-channel
\uboson~boson production.

Using a sample of KLOE data collected during 2004--2005, corresponding 
to an integrated luminosity of 1.54~fb$^{-1}$, we derived a new limit on 
the kinetic mixing parameter, $\varepsilon^2$, approaching the dielectron mass threshold.

\section{KLOE detector}
\label{sec:kloe_detector}
The Frascati $\upphi$ factory, DA$\Phi$NE, is an \eplus{}\eminus{} collider 
running mainly at a center-of-mass energy of 1.0195~GeV, 
the mass of the $\upphi$ meson. 
Equal energy electron and positron beams
collide at an angle of $\sim$25~mrad, producing $\upphi$ mesons nearly 
at rest. 

\begin{figure}[htb]
\centering
\includegraphics[width=0.98\linewidth]{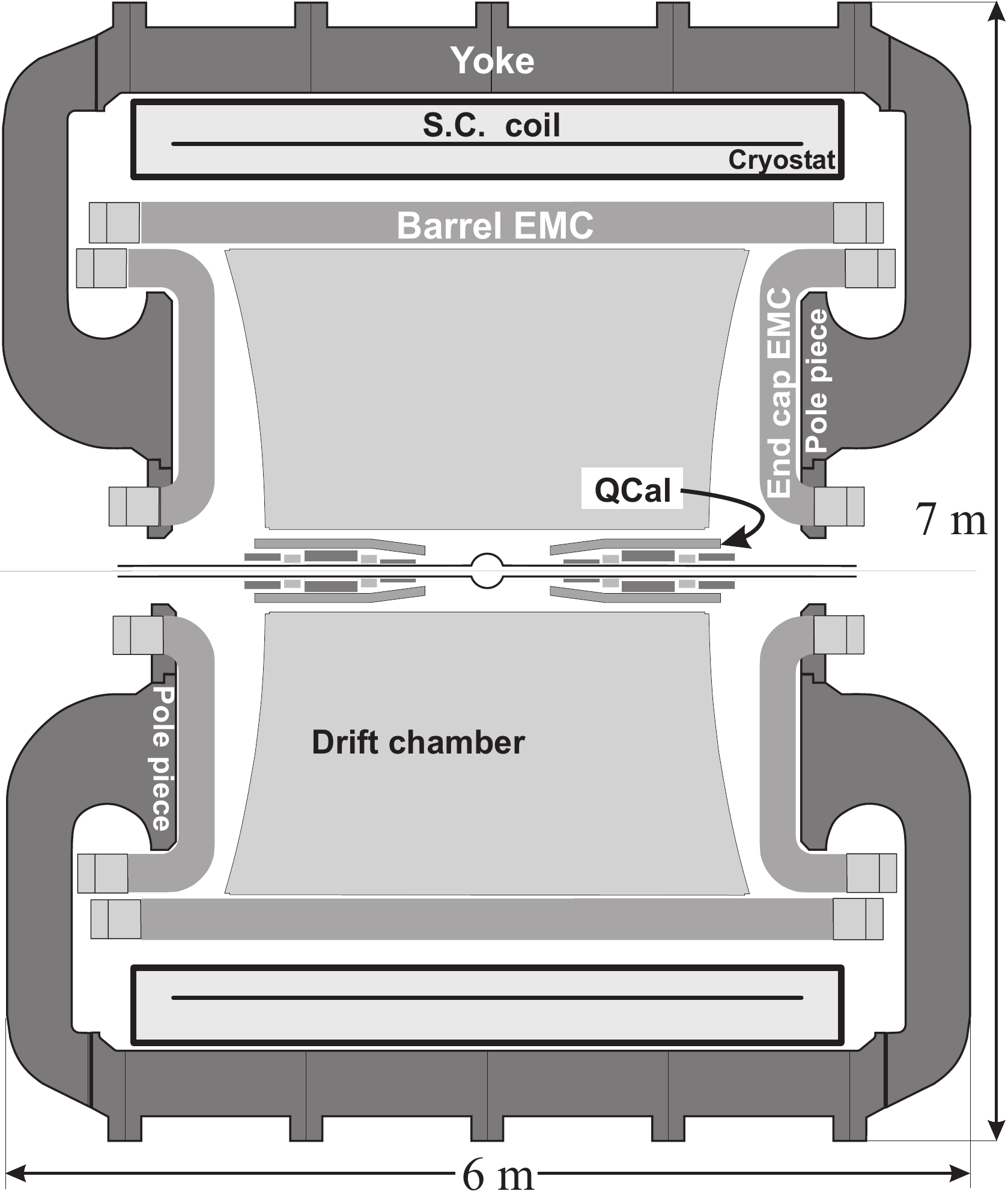}
\caption{Cross section of the KLOE detector.}
\label{Fig:KLOE_cross_section}
\end{figure}

%The KLOE experiment operated at DA$\Phi$NE from 2000 to 2006. 
The KLOE detector consists of a large cylindrical Drift Chamber (DC)~\cite{kloe_drift_chamber} 
with a 25~cm internal radius, 2~m outer radius, and 3.3~m length, comprising 
 $\sim$56,000 wires %of which $\sim$12,000 are stereo sense wires.
for a total of about 12,000 drift cells.
It is filled with a low-$Z$ (90\%~helium, 10\%~isobutane) gas mixture
and provides a momentum resolution of
$\sigma_{p_\bot}/p_\bot \approx$~0.4\%. The DC is surrounded by a
lead-scintillating fiber electromagnetic calorimeter (EMC)~\cite{kloe_calorimeter} 
composed of a cylindrical barrel and two end-caps providing 98\% coverage of the total solid angle.
Calorimeter modules are read out at both ends by 4880 photomultiplier tubes,
ultimately resulting in an energy resolution of $\sigma_E/E = 5.7\%/\sqrt{E (\mathrm{GeV})}$  
and a time resolution of $\sigma_t$ = 57~$\mathrm{ps}/\sqrt{E (\mathrm{GeV})} \, \oplus$~100~ps.
A superconducting coil around the EMC provides a 0.52~T field to measure the momentum of charged
particles. 
A cross sectional diagram of the KLOE detector is shown in Figure~\ref{Fig:KLOE_cross_section}.

The trigger~\cite{kloe_trigger_2002} uses energy deposition in the calorimeter
and drift chamber hit multiplicity. 
To minimize backgrounds the trigger system includes a second-level cosmic-ray muon veto 
based on energy deposition in the outermost layers of the calorimeter, followed by a 
software background filter based on the topology and
multiplicity of energy clusters and drift chamber hits to reduce beam background. 
A downscaled sample is retained to evaluate the filter efficiency.

\section{Event selection}\label{sec:selection}
Using 1.54~fb$^{-1}$ of KLOE data %collected during 2004--2005 
we have searched for 
\uboson~boson production in the process $\eplus\eminus \to \uboson\photon$ followed
by $\uboson \to \eplus\eminus$.
The center-of-mass energy of the collision depends on the amount of energy carried 
away by the initial-state radiation (ISR) photon. %, i.e. the radiative-return process.
The irreducible background originates from the 
$\eplus\eminus \to \eplus\eminus\photon$ radiative Bhabha scattering 
process, having the same three final-state particles. 
The reducible backgrounds consist of $\eplus\eminus \to \muplus\muminus\photon$, 
$\eplus\eminus \to \piplus\piminus\photon$, $\eplus\eminus \to \photon\photon$ (where one photon 
converts into an $\eplus\eminus$ pair), 
and $\eplus\eminus \to \upphi \to \uprho\uppi^{0} \to \piplus\piminus\uppi^{0}$,
as well as other $\upphi$ decays.
The expected \uboson~boson signal would appear as a resonant peak 
in the invariant-mass distribution of the $\eplus\eminus$ pair, $m_{\mathrm{ee}}$. This search differs 
from the previous KLOE searches~\cite{kloe_phi_eta_1,kloe_phi_eta_2,kloe_mmg} in its capability to probe the low mass region close to the 
dielectron mass threshold.

\begin{figure}[htb]
\centering
\includegraphics[width=0.98\linewidth]{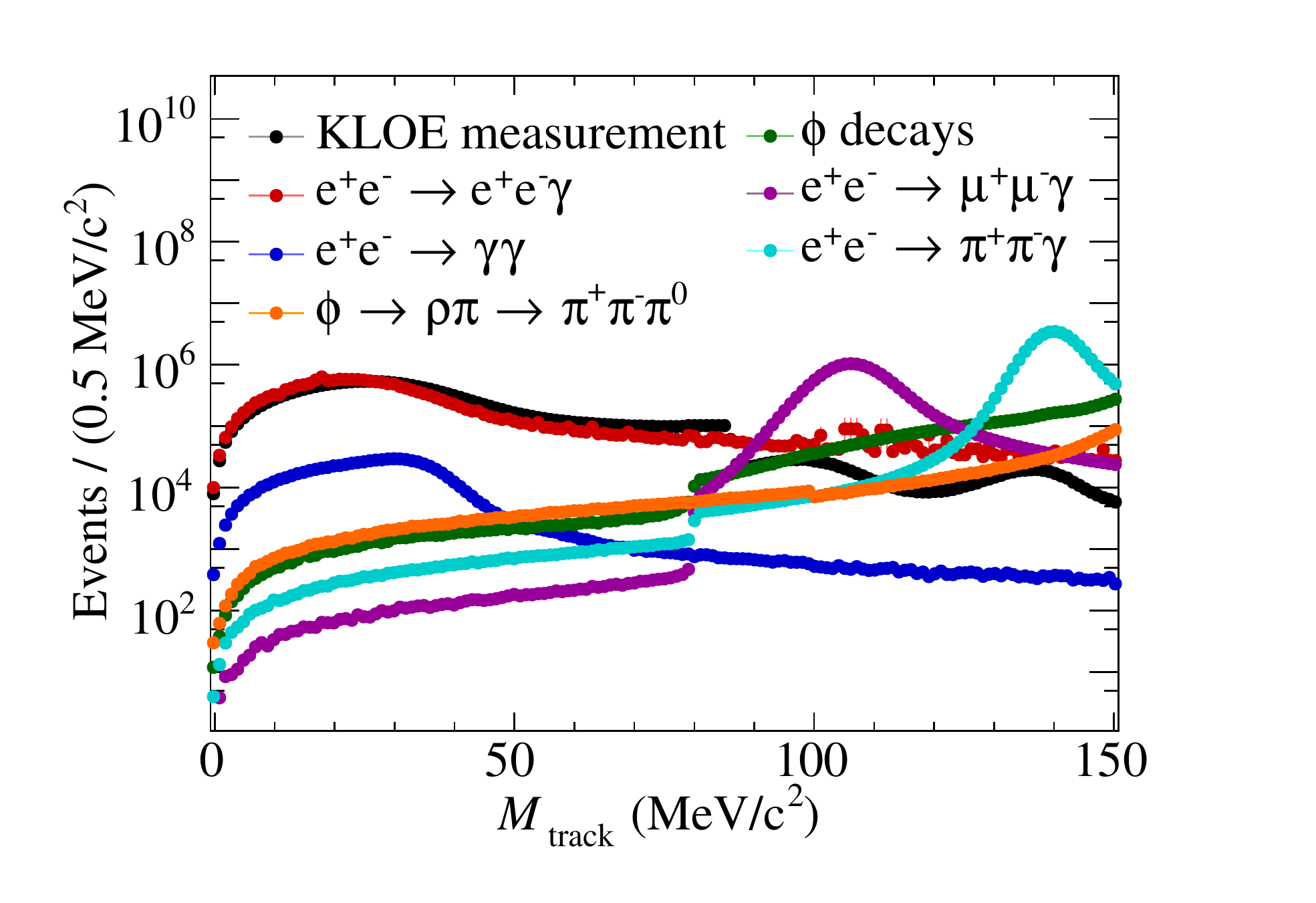}%{./figures/compare_final_Mtrk.pdf}
\caption{(color online) The track mass distribution before event selection for measurement and expected background simulations. 
Some of the simulations had a prescaling for $M_{\mathrm{track}}<80$~\MeVcc{}, which has been accounted for in 
the background evaluation. The measurement data set was prescaled for $M_{\mathrm{track}}>85$~\MeVcc{}. The track mass variable peaks at the mass of the charged track in the final state
for events with two charged tracks and a photon. The selection region is $M_{\mathrm{track}}<70$~\MeVcc{}.}
\label{Fig:track_mass}
\end{figure}

We selected events with three separate calorimeter energy deposits corresponding to two oppositely-charged lepton tracks
and a photon. The final-state electron, positron, and photon were required to be emitted at
large angle ($55^{\circ}<\theta<125^{\circ}$) with respect to the beam axis,
such that they are explicitly detected in the barrel of
the calorimeter, see Figure~\ref{Fig:KLOE_cross_section}. 
The large-angle selection greatly suppresses the $t$-channel contribution from the irreducible Bhabha-scattering background 
which is strongly peaked at small angle.
Since we are interested mostly
in the low invariant-mass region, we select only events with a hard photon, $E_{\photon}>305$~MeV, chosen 
to select a subsample of the events generated by our MC simulation.
We required both lepton tracks to have a first DC hit within a radius of 50~cm from the 
beam axis and a point-of-closest-approach (PCA) to the beam axis within the fiducial cylinder, $\rho_{{}_{\mathrm{PCA}}}<1$~cm 
and $-6<z_{{}_\mathrm{PCA}}<6$~cm, entirely contained within the vacuum pipe eliminating background events from photons converting on the vacuum wall.
We eliminated tightly spiralling tracks by requiring 
either a large transverse or a large longitudinal momentum for each of the lepton tracks, 
$p_{{}_\mathrm{T}}>160$~MeV/c or $p_z>90$~MeV/c. 
We require that the total momentum of 
the charged tracks is $\left( \left| p_{\eplus} \right| + \left| p_{\eminus} \right| \right) > 150$~\MeVc{} to avoid the presence of poorly reconstructed tracks.
A pseudo-likelihood discriminant was used to separate electrons from muons and pions~\cite{denig_kloe_note}.
A further discrimination from muons and pions was achieved using the $M_{\mathrm{track}}$ variable.
$M_{\mathrm{track}}$ is the $X$ mass for an $X^{+}X^{-}\photon$ final state, computed using energy and momentum conservation, 
assuming $m_{X^{+}} = m_{X^{-}}$~\cite{denig_kloe_note}.
In Figure~\ref{Fig:track_mass} the $M_{\mathrm{track}}$ distribution is reported for measured data and for all 
the relevant MC simulated background components. Including the cut $M_{\mathrm{track}}<70$~\MeVcc{} we were left
with 681,196 events at the end of the full analysis chain.

\section{Simulation and efficiencies}\label{sec:efficiency} %Systematic errors and efficiencies
\begin{figure}[htb]
\centering
\includegraphics[width=\linewidth]{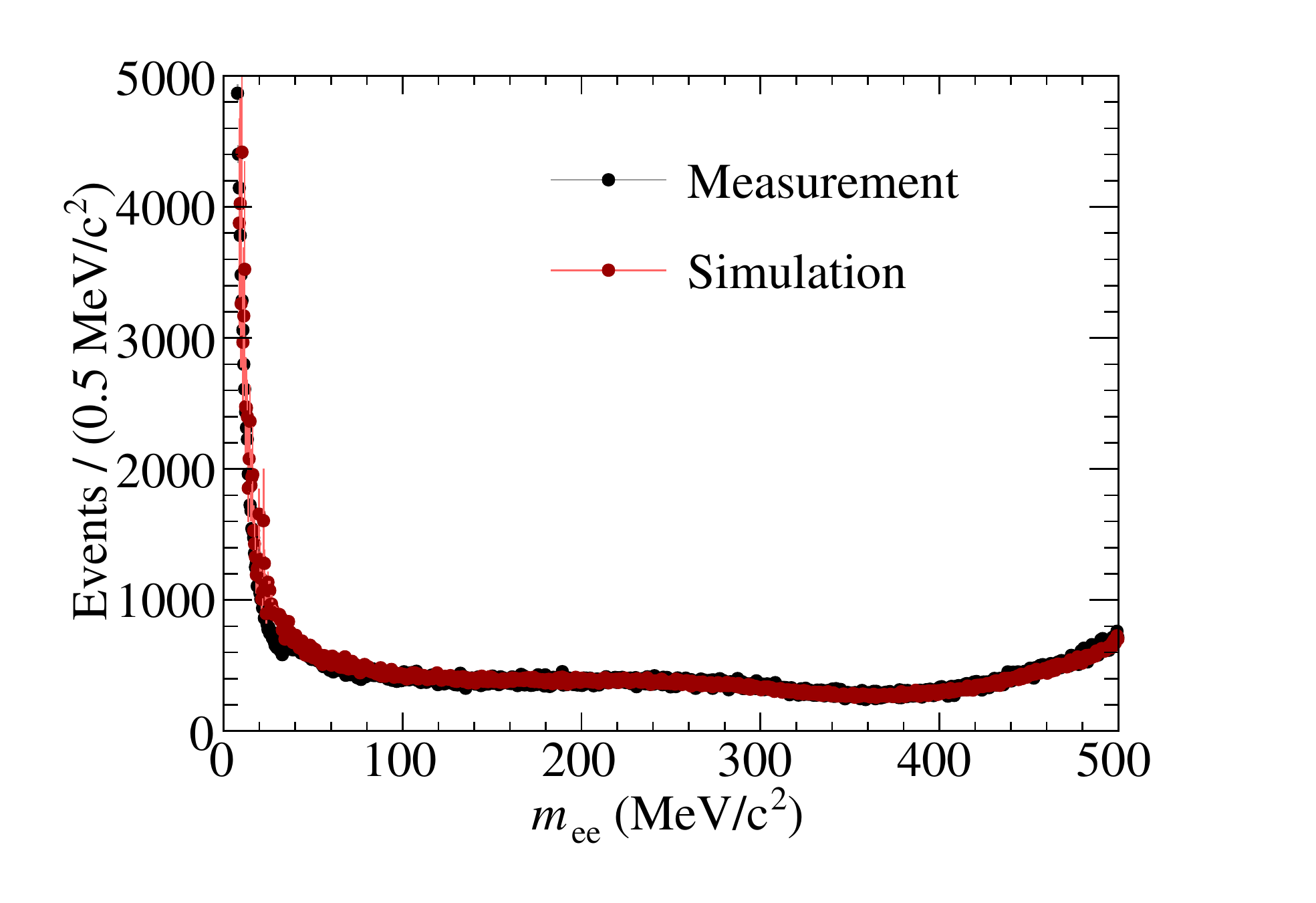}
\caption{(color online) Dielectron invariant-mass distribution from 
   measurement data with non-irreducible backgrounds subtracted compared to the {\sc Babayaga}-NLO MC simulation.}
\label{Fig:m_ee_distribution}
\end{figure}

We used MC event generators interfaced with the full KLOE detector simulation, GEANFI~\cite{kloe_eventClassification_2004}, including 
detector resolutions and beam conditions on a run-by-run basis,
to estimate the level of background contamination 
due to all of the processes listed in the previous section.
%due to the following processes: $\eplus\eminus \to \eplus\eminus\photon$, $\eplus\eminus \to \muplus\muminus\photon$, 
%$\eplus\eminus \to \piplus\piminus\photon$, $\eplus\eminus \to \photon\photon$ (where one photon 
%converts into an $\eplus\eminus$ pair), 
%and $\eplus\eminus \to \upphi \to \uprho\uppi^{0} \to \piplus\piminus\uppi^{0}$,
%as well as other $\upphi$ decays.
Excluding the irreducible background from radiative Bhabha scattering events, 
the contamination from the sum of residual backgrounds after all analysis cuts is less than
%the sum of the residual background contamination after all analysis cuts is less than
1.5\% in the whole $m_{\mathrm{ee}}$ range,
and none of the background shapes are peaked, 
eliminating the possibility of a background mimicking the resonant \uboson~boson signal.
The irreducible Bhabha scattering background was simulated using
the {\sc Babayaga}-NLO~\cite{babayaga1,babayaga2,babayaga3,babayaga4} event generator implemented within GEANFI 
(including the $s$-, $t$-, and $s$-$t$ interference channels) and is shown in
Figure~\ref{Fig:m_ee_distribution} along with the measured data after subtracting the non-irreducible background
components. No signal peak is observed.

\begin{figure}[htb]
\centering
\includegraphics[width=0.98\linewidth]{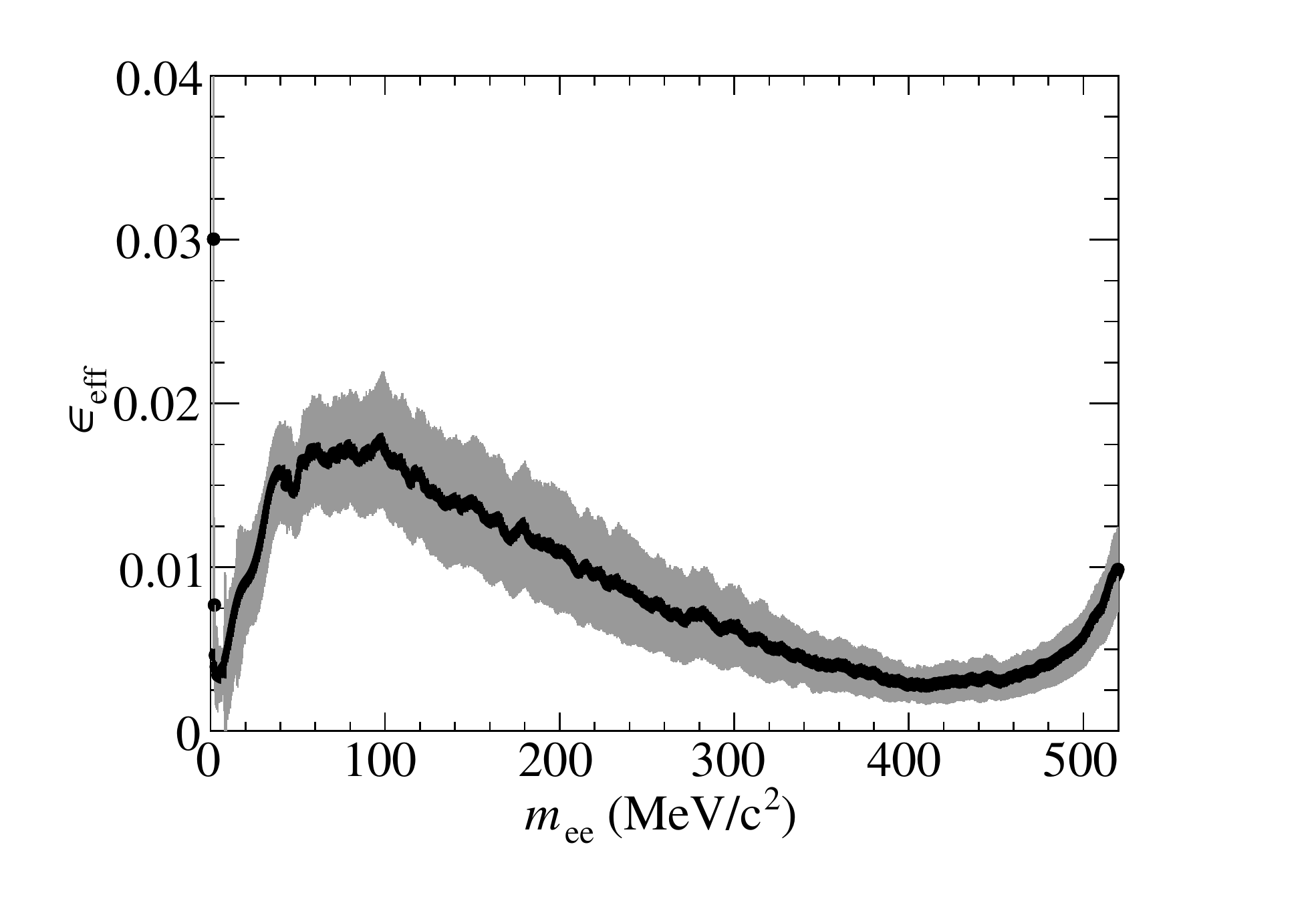}
\caption{Smoothed distribution of the total efficiency defined as the product of the selection efficiency for the 
$\eplus\eminus \to \uboson\photon \to \eplus\eminus\photon$ final state
evaluated using the Babayaga-NLO event generator modified to allow only the $s$-channel process,
and the generator-level efficiency evaluated from a {\sc Phokhara} MC simulation.}
\label{Fig:efficiency}
\end{figure}

In order to evaluate the \uboson~boson selection efficiency we used a modified version 
of the {\sc Babayaga}-NLO event generator implemented within GEANFI, such that
the radiative Bhabha scattering process was only allowed
to proceed via the annihilation channel, %since that is the channel 
in which the \uboson~boson resonance would occur. %; the $t$-channel ultimately becoming a background.
%The software background filter efficiency was determined directly from a subset of measurement data for which the filter 
%was not applied.
In order to create a large-statistics sample in our region of interest we restricted the {\sc Babayaga}-NLO generated
events to within $50^{\circ}<\theta_{\eplus,\eminus}^{\mathrm{MC}}<130^{\circ}$ and $E_{\photon}^{\mathrm{MC}}>300$~MeV.
The generator-level efficiency due to this restriction 
%($E_{\photon}^{\mathrm{MC}}>300$~MeV, $50^{\circ}<\theta_{\eplus,\eminus}^{\mathrm{MC}}<130^{\circ}$) 
was evaluated using a {\sc Phokhara} MC simulation~\cite{PHOKHARA}.
%The total efficiency is evaluated as the product of the software filter, the selection efficiency, and the geometrical and generator-level efficiency
%and is shown in Figure~\ref{Fig:efficiency}. 
The total efficiency is evaluated as the product of 
the generator-level efficiency 
and the event-selection efficiency, containing the cuts in Section~\ref{sec:selection} conditioned to the generator-level restriction as well 
as the trigger efficiency, 
and is shown in Figure~\ref{Fig:efficiency}. 
The decrease in efficiency as $m_{\electron{}\electron{}} \to 2m_{\electron{}}$ 
comes from the requirement on the total momentum of the charged tracks.

\section{Upper limit evaluation}%Upper limit on dark photon coupling
\label{sec:upper_limit}

\begin{figure}[htb]
\centering
\includegraphics[width=\linewidth]{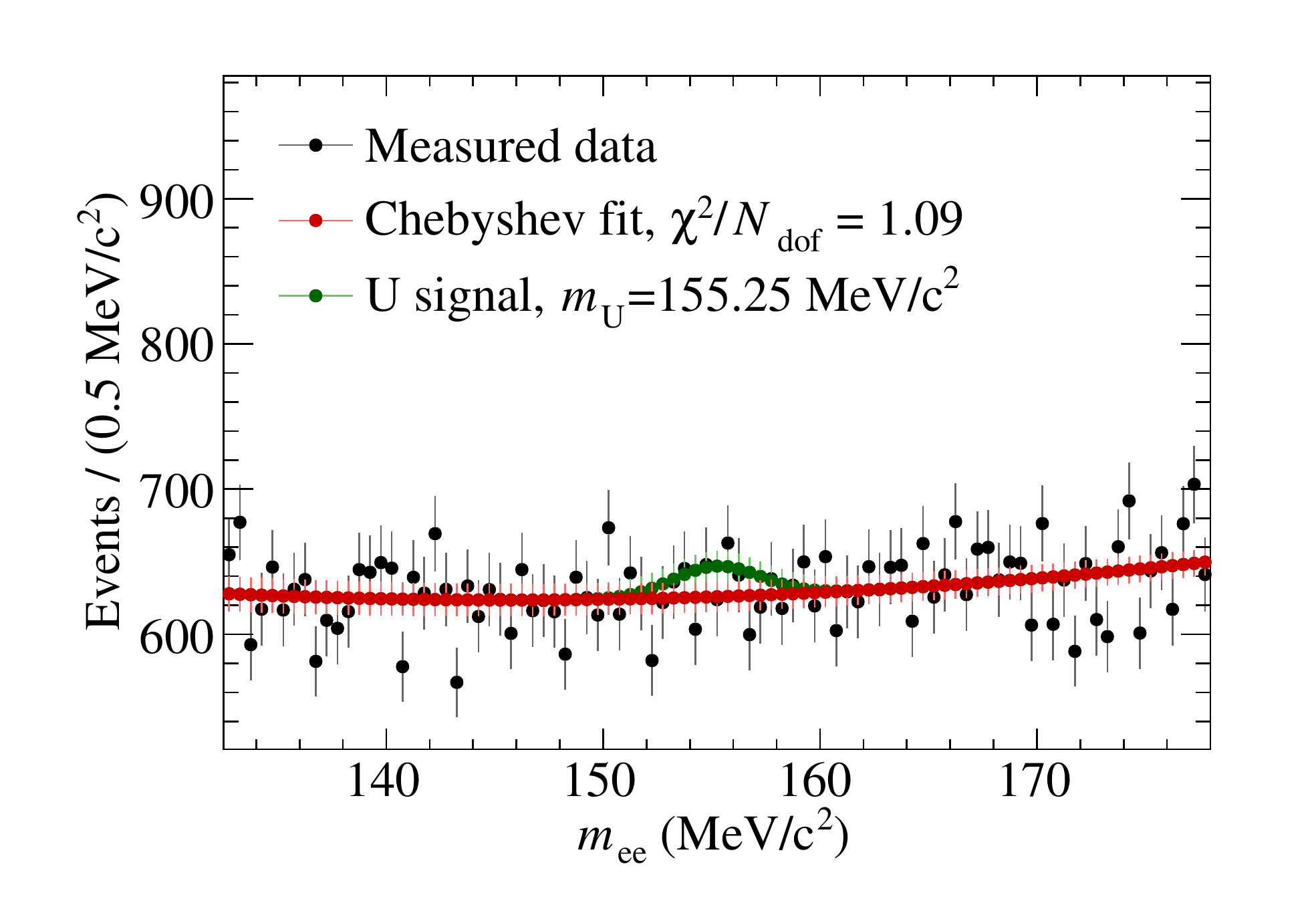}
\caption{(color online) The CL$_\mathrm{S}$ result at 90\% CL for $m_{\uboson} = 155.25$~MeV/c$^{2}$ showing 
the measured data, the Chebyshev-polynomial sideband 
fit, and the signal shape scaled to the CL$_\mathrm{S}$ result.}
\label{Fig:cls_result_example}
%\end{minipage}
\end{figure}

We used the CL$_\mathrm{S}$  technique~\cite{cls} to determine the limit on the number of signal
\uboson~boson events, $N_{\uboson}$, at 90\% confidence level using the $m_{\electron\electron}$ distribution. 
The invariant-mass resolution, $\sigma_{m_{\electron\electron}}$, is in the range $1.4 < \sigma_{m_{\electron\electron}} < 1.7$~\MeVcc{}.
Chebyshev polynomials were fit to the measured data ($\pm$15$\sigma_{m_{\electron\electron}}$), 
excluding the signal region of interest ($\pm$3$\sigma_{m_{\electron\electron}}$). The polynomial with $\chi^2/N_{\mathrm{dof}}$ closest 
to 1.0 was used as the background. 
A Breit-Wigner peak with a width of 1~keV smeared with the invariant-mass resolution was
used as the signal. An example of one specific CL$_\mathrm{S}$ result is shown in Figure~\ref{Fig:cls_result_example},
yielding an upper limit of $N_{\uboson} = 215$ \uboson~boson events 
at $m_{\uboson} = 155.25$~MeV/c$^2$ %that cannot be excluded 
at the 90\% confidence level. The $\chi^2/N_{\mathrm{dof}}$ was 1.09 for this Chebyshev-polynomial sideband fit.

\begin{figure}[htb]
\centering
\includegraphics[width=0.98\linewidth]{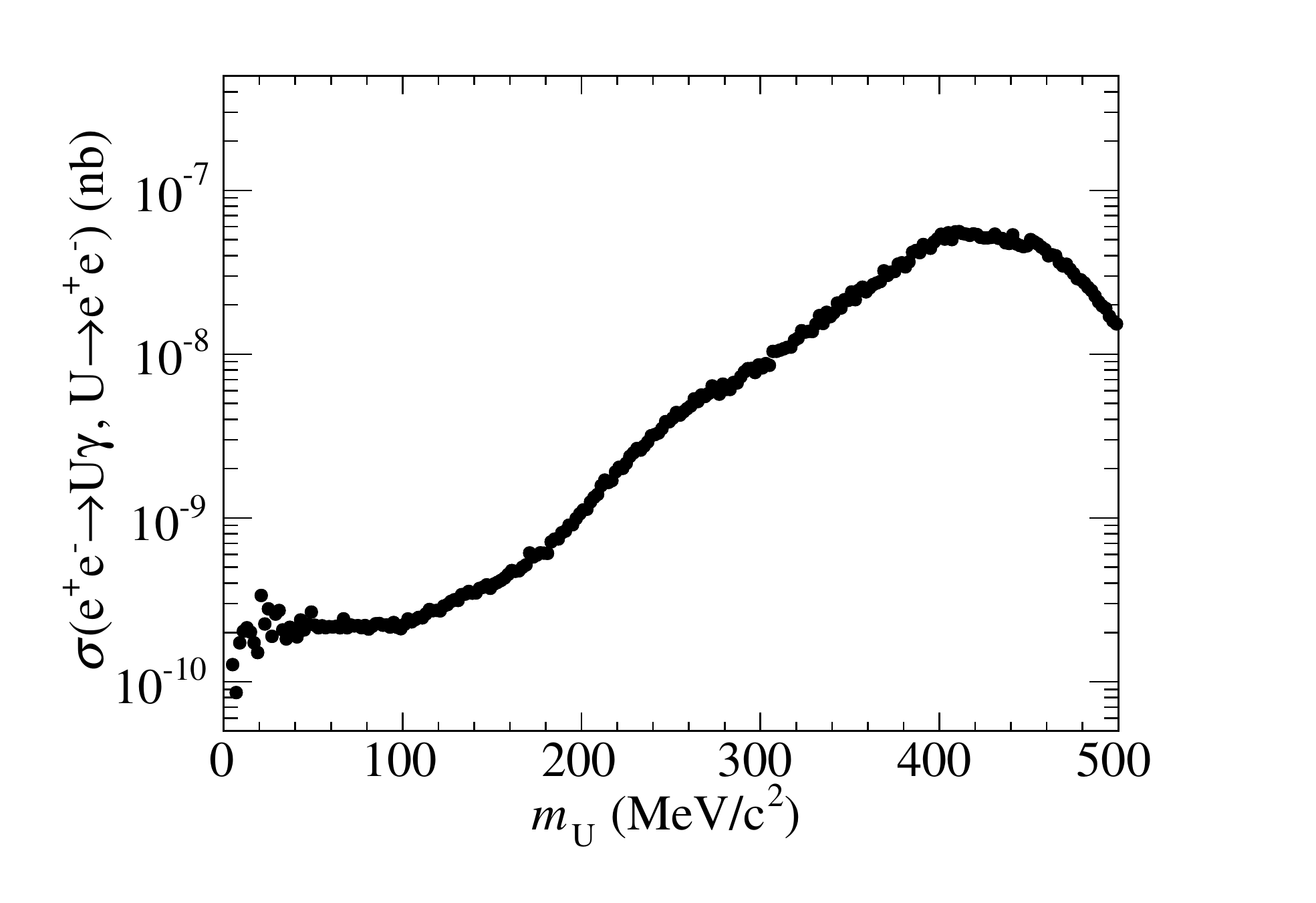}
\caption{Upper limit on the cross section 
$\sigma\left( \eplus\eminus \to \uboson\photon, \uboson \to \eplus\eminus\right)$.}
\label{Fig:cross_section_limit}
\end{figure}

%\section{Upper limit on the cross section}
% you just need to apply the formula that goes from x-sec to number of
% events at the end of the analysis chain:
% N(ee_Ug->eeg) = x-sec(ee_Ug->eeg) * Lumi * effi(ee_Ug->eeg)
% where effi(...) is the total analysis efficiency for signal events.
% This means that you U.L. on N from CLs directly translates in an U.L.
% on the x-sec:
% UL[x-sec(ee_Ug->eeg)] = UL[N(ee_Ug->eeg)] / (Lumi*effi(ee_Ug->eeg))
The upper limit at 90\% confidence level on the number of \uboson~boson
events, $\mathrm{UL}\left(N_\uboson\right)$, can be translated into a limit 
on the cross section,
%Our limit on the number of excluded \uboson~boson events can be translated into
%a 90\% confidence level limit on the cross section,
\begin{equation}
\mathrm{UL}\left[\sigma\left(\eplus\eminus \to \uboson\photon, \uboson\to\eplus\eminus\right)\right] = \frac{\mathrm{UL}\left(N_\uboson\right) }{ L \,\, \epsilon_{\,\mathrm{eff}}} \,\,,   
\end{equation}
%where $\mathrm{UL}\left(N_\uboson\right)$ is the number of possible \uboson~boson events we cannot exclude at
%the 90\% confidence level, 
where $L$ is the luminosity and $\epsilon_{\mathrm{eff}}$ is the total selection
efficiency. The limit is shown in Figure~\ref{Fig:cross_section_limit}.

\begin{figure}[h]
\centering
\includegraphics[width=0.99\linewidth]{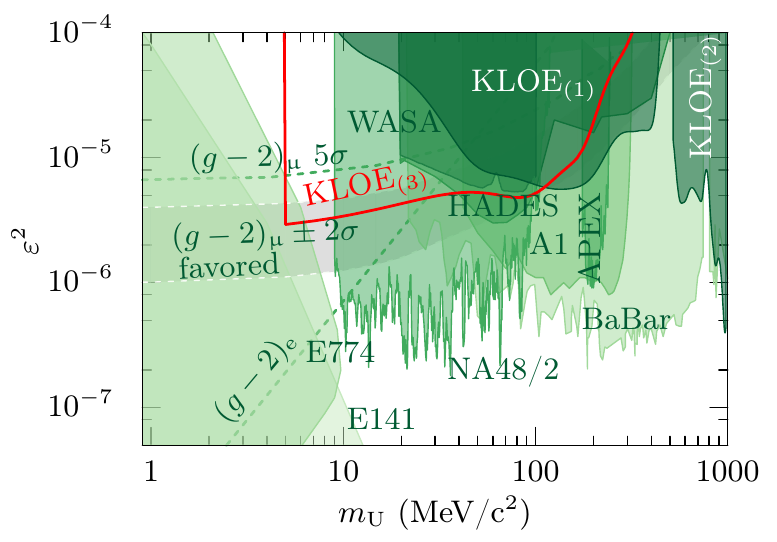}
\caption{(color online) Exclusion limits on the kinetic mixing parameter squared, $\varepsilon^2$, as a function
of the \uboson~boson mass. The red curve labeled KLOE$_{(3)}$ is the result of this article 
while the curves labeled KLOE$_{(1)}$ and KLOE$_{(2)}$ indicate the previous KLOE results. Also shown are the exclusion 
limits provided by E141, E774, Apex, WASA, HADES, A1, BaBar, and NA48/2. 
The gray band delimited by the dashed white lines indicates the mixing level and $m_\uboson$ parameter space that could explain the 
discrepancy observed between the measurement and SM calculation of the muon $(g\!-\!2)_{\upmu}$.}
\label{Fig:limits}
\end{figure}

We then translated the limit on $N_{\uboson}$ to a 90\% confidence level limit on the kinetic mixing parameter 
as a function of $m_{\mathrm{ee}}$ as in~\cite{kloe_mmg},
\begin{equation}
\label{eq:epsilon2}
\varepsilon^2\!\left(m_{\mathrm{ee}}\right) = \frac{ N_{\uboson}\!\left(m_{\mathrm{ee}}\right) }{\epsilon_{\,\mathrm{eff}}\!\left(m_{\mathrm{ee}}\right) }
                                   \frac{1}{ H\!\left(m_{\mathrm{ee}}\right) \, I\!\left(m_{\mathrm{ee}}\right) \, L } \,\,\, ,
\end{equation}
where the radiator function $H\!\left(m_{\mathrm{ee}}\right)$ was extracted from 
\begin{equation*}
\mathrm{d}\sigma_{\mathrm{ee}\photon}/\mathrm{d}m_{\mathrm{ee}} = H\!\left(m_{\electron\electron},s,\mathrm{cos}(\theta_\photon)\right) \cdot \sigma^{\mathrm{QED}}_{\mathrm{ee}}\!\left(m_{\mathrm{ee}}\right)
\end{equation*}
using the {\sc Phokhara} MC simulation~\cite{PHOKHARA} to determine the radiative
differential cross section, $I\!\left(m_{\mathrm{ee}}\right)$ is the integral of 
the cross section $\sigma(\eplus\eminus \to \uboson \to \eplus\eminus)$,
$L = 1.54$~fb$^{-1}$ is the integrated luminosity, and $\epsilon_{\mathrm{eff}}\left(m_{\mathrm{ee}}\right)$ is 
the total efficiency described in Section~\ref{sec:efficiency}.
Our limit is shown in Figure~\ref{Fig:limits} along with the indirect limits from the measurements 
of $\left( g-2 \right)_\electron$ and $\left( g-2 \right)_\upmu$ at 5$\sigma$ shown with dashed curves.
Limits from the following direct searches are shown with shaded regions and solid curves:
%E141~\cite{e141_1987},
%E774~\cite{e774_1991}, 
E141~\cite{dark_e141_e774},
E774~\cite{dark_e141_e774}, 
KLOE($\upphi \to \upeta \uboson$, $\uboson \to \eplus\eminus$)~\cite{kloe_phi_eta_1,kloe_phi_eta_2}, 
Apex~\cite{Apex2011},        
WASA~\cite{wasa2013},
HADES~\cite{hades_2014},
A1~\cite{A1_2014},
KLOE($\eplus\eminus \to \uboson\photon$, $\uboson \to \muplus\muminus$)~\cite{kloe_mmg},
BaBar~\cite{babar2014},
and NA48/2~\cite{na48_2_2015}.

\section{Systematic uncertainties}      

The background was determined by Chebyshev-polynomial sideband fits. The parameters of the polynomials were then
varied within 1$\sigma$ to determine the maximum variation of the polynomial shape. The uncertainty of each 
bin was set to the extent of that variation evaluated at the bin center. 
An example of the error bars on the Chebyshev-polynomial sideband fits can be seen in Figure~\ref{Fig:cls_result_example}. 
%The uncertainty in the background fit is therefore dominated by the statistical uncertainty in the selected data sample.
These bin uncertainties were taken into account in the CL$_\mathrm{S}$ procedure when %The TLimit procedure was instructed to take the bin uncertainties into account when 
determining $N_{\mathrm{CL_S}}\!\left(m_{\mathrm{ee}}\right)$. 
Since the irreducible background is smooth for each fit range,
we assume the Chebyshev polynomials sufficiently represent the background with negligible systematic uncertainty. 
Any uncertainty in the shape of the smeared resonant peak was also taken to be negligible.

The efficiency of the $\eplus\eminus\to\eplus\eminus\photon$ event selection was determined by 
taking the ratio of the set of simulated events that passed 
the selection criteria to the total simulated sample. We apply a 0.1\% systematic uncertainty due to 
the {\sc Babayaga}-NLO event generator~\cite{babayaga1,babayaga2,babayaga3,babayaga4}, a 0.1\% systematic uncertainty for the 
trigger, and a 0.1\% systematic uncertainty for the software background filter.
All together the uncertainty on the selection efficiency is dominated 
by the statistical uncertainty on the selected sample. 
A {\sc Phokhara} MC simulation~\cite{PHOKHARA} was performed
to evaluate the generator-level efficiency due to the restriction $E_{\photon}^{\mathrm{MC}}>300$~MeV and $50^{\circ}<\theta_{\eplus,\eminus}^{\mathrm{MC}}<130^{\circ}$. 
The selection efficiency and the generator-level efficiency are combined
to give the total efficiency, $\epsilon_{\,\mathrm{eff}}\!\left(m_{\mathrm{ee}}\right)$. The uncertainty is given as the error band in 
Figure~\ref{Fig:efficiency}, again dominated by the statistical uncertainties in the simulated data set.

There are two effects that
contribute to the uncertainty in the radiator function, $H\!\left(m_{\mathrm{ee}}\right)$. 
First, since the value of $H\!\left(m_{\mathrm{ee}}\right)$ is taken from
simulated data, we must take into account the statistical uncertainty on those
values. Second, we assume a uniform 0.5\% systematic uncertainty in the calculation
of $H\!\left(m_{\mathrm{ee}}\right)$, as quoted in \cite{radiator1,radiator2,radiator3,PHOKHARA}.
The uncertainty in the integrated luminosity is 0.3\%~\cite{denig_kloe_note}. %0.107\%.\\
The uncertainties on $H\!\left(m_{\mathrm{ee}}\right)$, $\epsilon_{\,\mathrm{eff}}\!\left(m_{\mathrm{ee}}\right)$,
and $L$, propagate to the systematic uncertainty on $\varepsilon^2\!\left(m_{\mathrm{ee}}\right)$ via~(\ref{eq:epsilon2}).
A summary of systematic uncertainties is presented in Table~\ref{tab:1}.

%Note: the errors may seem large, because we had weighted events which leads to large statistical uncertainties.

\begin{table}[h]
\centering
\caption{Summary of systematic uncertainties. The uncertainties on the efficiency, radiator function, and cross-section 
integral vary as a function of $m_{\electron\electron}$. The numbers quoted here correspond to the largest estimate within our $m_{\electron\electron}$ range.}\label{tab:1}
\begin{tabular}{l c} 
 Systematic source                  & Relative uncertainty \\
 \hline \\[-1.9ex]
Background (sideband fit)                & negl.        \\ 
 $\epsilon_{\mathrm{eff}}\left( m_{\electron\electron} \right)$ & 2\% \\
{\hspace{2.0ex} MC generator, 0.1\%}  & {} \\
{\hspace{2.0ex} Trigger, 0.1\%}  & {} \\
{\hspace{2.0ex} Software background filter, 0.1\%}  & {} \\
{\hspace{2.0ex} Event selection, 2\%}  & {} \\
 $H\left( m_{\electron\electron} \right)$  & 0.5\%  \\
 $I\left( m_{\electron\electron} \right)$  & negl. \\
 $L$ & 0.3\% \\  [1ex]
 \hline
 \end{tabular}
\end{table}

\section{Conclusions}
\label{sec:conclusions}
We performed a search for a dark gauge \uboson~boson in the 
process $\eplus\eminus \to \uboson\photon$ with $\uboson \to \eplus\eminus$ 
using the radiative return method and 1.54~fb$^{-1}$ of KLOE data 
collected in 2004--2005. We found no evidence for a \uboson~boson 
resonant peak and set a 90\% CL upper limit on the kinetic mixing 
parameter, $\varepsilon^2$, at $10^{-6}$--$10^{-4}$ in 
the \uboson-boson mass range 5--520~MeV/c$^2$. 
This limit partly excludes some of the remaining parameter space 
in the low dielectron mass region allowed by the discrepancy between the observed and predicted (g-2)$_\muon$.
%favored by the (g-2)$_\muon$  anomaly.

\section{Acknowledgments}
\label{sec:acknowledgments}

We warmly thank our former KLOE colleagues for the access to the data collected during the KLOE data taking campaign.
We thank the DA$\Phi$NE team for their efforts in maintaining low background running conditions and their 
collaboration during all data taking. We want to thank our technical staff: 
G.F.~Fortugno and F.~Sborzacchi for their dedication in ensuring efficient operation of the KLOE computing facilities; 
M.~Anelli for his continuous attention to the gas system and detector safety; 
A.~Balla, M.~Gatta, G.~Corradi and G.~Papalino for electronics maintenance; 
M.~Santoni, G.~Paoluzzi and R.~Rosellini for general detector support; 
C.~Piscitelli for his help during major maintenance periods. 
This work was supported in part by the EU Integrated Infrastructure Initiative Hadron Physics Project under 
contract number RII3-CT- 2004-506078; by the European Commission under the 7$^{\mathrm{th}}$ Framework Programme 
through the `Research Infrastructures' action of the `Capacities' Programme, Call: FP7-INFRASTRUCTURES-2008-1, 
Grant Agreement No. 227431; by the Polish National Science Centre through the Grants No. 
%0469/B/H03/2009/37, 
%0309/B/H03/2011/40, 
DEC-2011/03/N/ST2/02641, 
%2011/01/D/ST2/00748,
2011/03/N/ST2/02652,
2013/08/M/ST2/00323,
2013/11/B/ST2/04245,
2014/14/E/ST2/00262,
and by the Foundation for Polish Science through the MPD programme.
%and the project HOMING PLUS BIS/2011-4/3.

In addition, we would like to thank the {\sc Babayaga} authors,
C.M.~Carloni~Calame, G.~Montagna, O.~Nicrosini, and F.~Piccinini, for 
numerous useful discussions and help while modifying their code for our purpose.

\end{document}